\documentclass[12pt]{article}
\usepackage{amsfonts}
\usepackage{amssymb}

\makeatletter

\@addtoreset{equation}{section}
\def\section{\@startsection {section}{1}{\z@}{-3.0ex plus -1ex minus
 -.2ex}{2.3ex plus .2ex}{\large\bf}}
\def\subsection{\@startsection{subsection}{2}{\z@}{-2.25ex plus%
 -1ex minus -.2ex}{1.5ex plus .2ex}{\bf}}

\advance \voffset by -1.0in
\advance \hoffset by -0.6in
\def\halmos{\hbox{\vrule height0.31cm width0.01cm\vbox{\hrule height
 0.01cm width0.3cm \vskip0.29cm \hrule height 0.01cm width0.3cm}\vrule
 height0.31cm width 0.01cm}}

\def\cd{\!\cdot\!}

\def\bx{{\mbox{\boldmath $x$}}}

\def\bl{{\mbox{\boldmath $l$}}}
\def\bk{{\mbox{\boldmath $k$}}}

\def\bd{{\mbox{\boldmath $d$}}}
\def\Li{L^{(1)}_i}
\def\Lj{L^{(1)}_j}
\def\R{R^{(2)}_i}
\def\Rj{R^{(2)}_j}
\def\LLbar{\tilde{ L}^{(1)}_i}
\def\Rbar{\tilde{ R}^{(2)}_i}

\def\rinv{s}

\newcommand{\ZZ}{\mathbb{Z}}

\newcommand{\RR}{\mathbb{R}}

\newcommand{\tr}{{\rm tr}}
\def\bea{\begin{eqnarray}}
\def\eea{\end{eqnarray}}
\def\S{Section }
\newtheorem{theorem}{Theorem}[section]
\newtheorem{lemma}[theorem]{Lemma}

\newcommand{\oR}{\overline{\RR^3}}
\newcommand{\be}{\begin{equation}}
\newcommand{\ee}{\end{equation}}
\newcommand{\p}{\partial}
\newcommand{\Diff}{\mbox{Diff}}

\newcommand{\cA}{{\cal A}}
\newcommand{\cO}{{\cal O}}

\begin{document}
\begin{flushright}
EMPG-02-22\\
DAMTP-2002-142\\
HWM-02-37\\
hep-th/0212075
\end{flushright}
\begin{center}
\baselineskip 24 pt
{\Large \bf The interaction energy of well-separated }

{\Large \bf Skyrme solitons  }

\baselineskip 16pt
\parskip 7pt

\vspace{0.5 cm}
 N.~S.~Manton,
Centre for Mathematical Sciences, University of Cambridge, \\
       Wilberforce Road,  Cambridge CB3 0WA, UK\\
e-mail {\tt  N.S.Manton@damtp.cam.ac.uk} \\
\vspace{0.2cm}

 B.~J.~Schroers,
Department of Mathematics, Heriot-Watt University, \\
Riccarton, Edinburgh EH14 4AS, UK  \\
e-mail {\tt bernd@ma.hw.ac.uk} \\
\vspace{0.2cm}

M.~A.~Singer,
School of Mathematics, University of Edinburgh,\\
Kings Buildings,  Edinburgh EH9 3JZ, UK \\
e-mail {\tt michael@maths.ed.ac.uk}
\vspace{0.3cm}

{   5 December     2002}

\end{center}

\begin{abstract}
\noindent
We prove that the  asymptotic field of a
 Skyrme soliton of any degree has a non-trivial
multipole expansion. It follows that every Skyrme soliton
has a well-defined leading multipole moment.
We derive an  expression  for the linear  interaction energy of
well-separated Skyrme solitons in terms of their
leading multipole moments. This  expression can
 always be made negative by suitable
rotations of one of the Skyrme solitons in space and iso-space.
We show that
the linear interaction energy dominates for large
separation if the
orders of the Skyrme solitons' multipole moments
differ by at most two. In that case there
are therefore always attractive forces between
the Skyrme solitons.
\end{abstract}

\section{Skyrme solitons}

The fundamental field of Skyrme's  theory \cite{Skyrme}
is a map
\bea  
U: \RR^3 \rightarrow SU(2).
\eea
We  denote points in $\RR^3$ by $x$ with coordinates $x_i$, $i=1,2,3$
and Euclidean length  $
r=|x|=\sqrt{x_1^2+x_2^2 +x_3^2}$ .
Sometimes we write $\hat x$ for the unit vector $x/r$.
It is often useful to parametrise $U$ in terms of the Pauli matrices
$\tau_1, \tau_2$ and $\tau_3$   and the triplet of pion fields $\pi_1,\pi_2$
and $\pi_2$ as
\bea
\label{pions}
U(x) =   \sigma (x) + i\pi_a(x)\tau_a,
\eea
where summation over the repeated index $a$ is implied and
the field $\sigma $ is determined by the constraint $\sigma^2 + \pi_1^2+
\pi_2^2+ \pi_3^2 = 1$.
In this introductory section we do not specify
the class of functions to which $U$ belongs. It is assumed to
be sufficiently smooth for all the following operations to make sense.

The Skyrme energy functional is best written in terms  of
the Lie-algebra valued currents
\bea
\label{Lc}
L_i =  U^{\dagger}\partial_iU
\eea
or
\bea
\label{Rc}
R_i = U\partial_iU^{\dagger},
\eea
where  $\partial_i=\partial/\partial x_i$.
It is
\bea
\label{skyrpot}
E[U] = -\int d^3x \, \left({1\over 2}\tr(L_i L_i) + {1 \over 16}
       \tr(\lbrack L_j,L_i\rbrack \lbrack L_j,L_i\rbrack ) \right).
\eea
The Euler-Lagrange equation for stationary points of this
functional is conveniently expressed
in terms of  the modified currents
\bea
\label{LLc}
\tilde{L}_i =  L_i -{1\over 4 }
\lbrack L_j,\lbrack L_j,L_i\rbrack \rbrack
\eea
and
\bea
\label{RRc}
\tilde{R}_i =  R_i - {1\over 4 }
\lbrack R_j,\lbrack R_j,R_i\rbrack \rbrack,
\eea
where we again use the convention that repeated indices are summed over.
It  reads
\bea
\label{ELL}
\partial_i \tilde{L}_i =   0
\eea
or, equivalently,
\bea
\label{ELR}
\partial_i \tilde{R}_i=0.
\eea

Here we are interested in finite-energy solutions of
 the Euler-Lagrange equation.
It is shown in \cite{Esteban} that
the finite energy requirement implies that the map $U$ tends to a
constant value at infinity in a weak sense.
We choose that constant to be  the identity
element $1\in SU(2)$ and demand
\bea
\label{boundary}
\lim_{r\rightarrow \infty} U(x)=1.
\eea
The boundary condition (\ref{boundary}) means that the domain of $U$
is  effectively compactified to a three-sphere.
Since the target space is also
a three-sphere, maps satisfying (\ref{boundary}) have an associated
integer degree. The first rigorous proof that for
 a finite-energy Skyrme configuration in a very general class of 
functions
the degree
\bea
\label{degreeint}
\mbox{deg}[U]=
-\frac{1}{24 \pi^2}\int d^3 x\, \epsilon_{ijk}\,\tr\left(L_iL_jL_k\right)
\eea
is an integer was given in  \cite{EM}.
This result means that the space
\bea
{\cal C}  =\{ U: \RR^3 \rightarrow SU(2)\,|\, E[U] < \infty \}
\eea
of finite-energy configurations is a disjoint union of sectors
\bea
{\cal C}_k=\{ U: \RR^3 \rightarrow SU(2)\, |\,E[U]
<\infty \quad \mbox{and \, deg}[U]=k \}
\eea
labelled by the integers $k\in \ZZ$.

The symmetry group of Skyrme's theory will play an important role
in our discussion.
The energy functional (\ref{skyrpot}), the boundary condition
(\ref{boundary}) and the degree (\ref{degreeint})
 are invariant under the action of the Euclidean
group of translations and rotations in $\RR^3$ and under
rotations of the pion fields
\bea
\pi_a \mapsto G_{ab}\pi_b, \qquad G\in SO(3),
\eea
which we call iso-rotations. Reflections in space
$S:  x\mapsto -x$
and   in iso-space $\pi_a\mapsto  -\pi_a$ both leave the energy
invariant but each changes the sign of the degree. The pull-back
of  Skyrme configurations $U$ via $S$ provides a map
\bea
\label{sectref}
\tilde S: {\cal C}_k\rightarrow {\cal C}_{-k}, \quad
U\mapsto U\circ S
\eea
which preserves the energy.

It was shown in \cite{Faddeev}
 that the energy in each topological sector  is bounded
below by  a multiple of the degree.
It follows from the results of \cite{Manton}
that the bound cannot be attained for the standard version
of the Skyrme model described here,
so that we have the strict inequality
\bea
\label{bbelow}
E[U] > 12\pi^2 |k|.
\eea
The bound ensures that the
infima
\bea
\label{infdef}
I_k=\mbox{inf}\{ E[U]\,|\,U\in {\cal C}_k\}
\eea
are well-defined. The question of whether the infima are attained
was first addressed by Esteban  in  the paper \cite{Esteban}.
 Amongst other things
Esteban proved that, for a suitable class of functions,
\bea
\label{winequality}
I_k\leq I_l+ I_{k-l}
\eea
for all $k,l\in \ZZ$.
She also showed that infima are attained
provided one assumes the strict inequality
\bea
\label{inequality}
I_k<I_l+ I_{k-l}
\eea
for all $k\in\ZZ-\{0,\pm 1\}$ and $l\in \ZZ-\{0,k\}$
in the range $|l| + |k-l|< \sqrt 2 |k|$.
In  \cite{EM} it was shown that the result still holds if one widens
the class of allowed functions
but the inequality (\ref{inequality}) remains a necessary assumption
in the proof.
The strict inequality is also of interest in
physics. As we shall see it is related to the question of
   attractive forces in the Skyrme model.

For low values of $k$ the existence and nature of minima of the
Skyrme energy functional is understood in more detail.
For fields of degree one, the highly symmetric hedgehog ansatz
\bea
\label{hedge}
U_H(x)=\exp (i f(r)\hat{x}_a\tau_a), 
\eea
 introduced already by Skyrme, leads to
an ordinary differential equation for the profile function $f$.
With the boundary conditions $f(0)=\pi$ and $f(\infty)=0$
the resulting Skyrme configuration has degree 1 and is called the
Skyrmion. It was shown in \cite{KL} that the Skyrmion minimises the
Skyrme energy functional amongst all degree one configurations of
the hedgehog form. In
 \cite{Esteban} the existence of the minimum
of the Skyrme energy functional in ${\cal C}_1$  was proved,
but it has not been established rigorously that  minimising configurations
have the symmetry of the hedgehog field (\ref{hedge}).
Note that if $U_1$ is a minimal energy configuration in ${\cal C}_1$
then the reflected configuration  $\tilde S \left( U_1\right)$
has the same energy and
minimises the  energy  in   ${\cal C}_{-1}$.

In the following we use the term Skyrme solitons for minimal
energy solutions of the Skyrme equation with non-zero degree $k$.
There is overwhelming numerical evidence that the minimum in
${\cal C}_2$ is attained by a configuration of toroidal symmetry
\cite{torus}. For higher degree, too, much is known numerically
about the minima of Skyrme's energy functional in the sector  ${\cal C}_k$.
Numerical searches, assisted by analytical ans\"atze and investigations
of the possible symmetries of Skyrme solitons,  suggest the
existence of Skyrme solitons in all sectors ${\cal C}_k$ up to
$k=22$. The energies are sufficiently accurately computed that it appears
that the inequality (\ref{inequality}) is satisfied for all $k$
in the range $2\leq k\leq 20$ and all $l$ in the range $0<l<k$, see
\cite{BCT,BS1,BS2}.

The existence of attractive forces between  two
Skyrmions was already shown by Skyrme, using the product ansatz.
In this paper we investigate the existence of attractive
forces between general Skyrme solitons.
 It is perhaps
worth stressing that our arguments in fact apply to any finite-energy
solution of the Skyrme equation, not just minimal ones.
An earlier attempt at proving the existence of attractive
forces between Skyrme solitons  was made in
the unpublished paper \cite{CK}.
Our approach  is partly inspired by
ideas in \cite{CK} but also fills important  gaps left there.
Our main tool is an asymptotic expansion of Skyrme solitons.
In Section 2 we show that  Skyrme fields have an
asymptotic expansion in powers of $1/r$ and that for non-trivial
Skyrme solitons that expansion always has a non-trivial leading multipole.
In Section 3 we study the interaction energy of two Skyrme solitons
and show that it is dominated by a  certain linear interaction  energy
provided the orders of the leading multipoles of the Skyrme solitons
do not differ by more than 2. In Section 4 we derive an expression
for  the  linear interaction energy  of two multipoles, and show that it can
always be made negative by  suitable relative rotations
in space and iso-space.
At the end of this paper we briefly comment on the relationship
between our results and Esteban's work, and on the implications
for the existence of Skyrme solitons of arbitrary degree.

\section{The asymptotics of Skyrme solitons}
\label{asymptotics}

The aim of this section is to show that if $U$ and the currents $L_j$
are a little better than continuous, then $U$ is smooth in $\RR^3$, and
has a non-trivial asymptotic expansion in powers of $1/r$ and $\log r$
as $r\to \infty$. By {\em non-trivial}  we mean here that if $U$
is non-constant, then it cannot happen that all terms in the
asymptotic expansion vanish.  Put another way, it is not possible for
$U$ to approach $1$ at infinity faster than {\em every} power of $1/r$ unless
$U$ is identically equal to $1$ in $\RR^3$.

To make a precise statement, say that the (possibly matrix-valued) 
function $f$ is in $ C_b^{0,\alpha}(\RR^3)$,
with  $0<\alpha<1$, if  $f$ satisfies 
\be\label{10.1.11}
 \sup_{x\in \RR^3} |f| + \sup_{x,x'\in \RR^3, x\neq x'}
\frac{(1+r+r')^\alpha|f(x)-f(x')|}
{|r-r'|^\alpha + (1+r+r')^\alpha d(\omega,\omega')^\alpha} <\infty.
\ee
Here $x = r\omega, x' = r'\omega'$, where $\omega$ is regarded as an
angular variable (unit vector) living on the unit $2$-sphere, and
$d(\omega,\omega')$ is the $2$-sphere distance between $\omega$ and
$\omega'$. If we consider the space $C_b^{0,\alpha}(K)$, where $K$ is
a bounded subset of $\RR^3$, then this is precisely the same as the
usual space of (bounded) H\"older-continuous functions, with H\"older
exponent $\alpha$.  However, $C_b^{0,\alpha}(\RR^3)$ is slightly
different from the usual space of bounded H\"older-continuous
functions on $\RR^3$.

The main result of this section can be summarized as follows:
\begin{theorem}
\label{asymps}
Suppose that for some $\delta>0$,
\be \label{1.1.11}
(1+r)^\delta (U-1)\in C_b^{0,\alpha},\;\;
(1+r)^\delta L_j\in C_b^{0,\alpha}.
\ee
Suppose further that  $U$ satisfies the Skyrme differential equations
(\ref{ELL}) in the sense of distributions.

Then $U$ is smooth in $\RR^3$ and
has a complete asymptotic expansion in powers of $1/r$ and
$\log r$, for large $r$. If $U$ is non-constant, then this expansion has a
      leading term which is harmonic,       hence a multipole.
\end{theorem}

Note that the hypotheses (\ref{1.1.11}) force $U$ to approach $1$ and
the $L_j$ to approach $0$ like $r^{-\delta}$ as $r\to \infty$.
As a technical remark, we point out that the
assumption of  H\"older-continuous currents implies that $U$
will have a H\"older continuous first derivative. The second
derivatives $\p_j\p_k U$ are then defined only in the sense of distributions,
but in the Skyrme equation $\p_j\p_k U$  enters linearly, and is
multiplied by continuous functions (smooth functions  of the currents)
see \S\ref{rewrite} below.  In particular, the left-hand
side of the Skyrme equation $\p_j\tilde{L}_j = 0$ makes sense as a
distribution if (\ref{1.1.11}) holds.

The assumptions (\ref{1.1.11}) of Theorem~(\ref{asymps})  do not
follow immediately from the variational analysis used by Esteban in
\cite{Esteban}. Her methods only give that the derivatives
$\p_j U$ are locally
square-integrable (and that the components of $U$ are locally
bounded).
On physical grounds, one expects  minimizers of the Skyrme energy
functional to satisfy (\ref{1.1.11}),  but it would be desirable to
bridge the gap between the analysis given here and what was proved
rigorously in \cite{Esteban} and \cite{EM}. This issue will not be
pursued further here.

The proof of Theorem~(\ref{asymps}) proceeds in four steps, each of which
takes up one of the following subsections.
In the first subsection we re-write the Skyrme equation in order to
make explicit the form of the non-linearities. The equation is {\em
quasilinear}, in the sense that the derivatives of highest order (2)
enter linearly. The Skyrme equation can therefore be regarded as a second-order
linear elliptic PDE, with H\"older-continuous coefficients. Then
standard regularity theorems (Schauder estimates) yield that $U$ is smooth.

In  \S\ref{conormal}
we study the behaviour of the Skyrme equation
at spatial infinity, by introducing coordinates
$(\rinv,\omega)$, where $\rinv = r^{-1}$, so that
the 2-sphere at infinity becomes a genuine boundary at
$\rinv=0$. After a rescaling, the Laplacian of $\RR^3$ is replaced by
\be \label{2.1.11}
\Delta_b = (\rinv\p_\rinv)^2 - \rinv\p_\rinv + \Delta_\omega,
\ee
where $\Delta_\omega$ is the Laplacian of the unit $2$-sphere. The
analysis is now guided by the corresponding analysis of a system of
ordinary differential equations 
 with regular singular point at $\rinv=0$.  In particular, one does
not expect the solutions to be smooth near $\rinv=0$, but one does
expect a non-trivial expansion in powers of $\rinv$ (and possibly $\log\rinv$).
The $b$-calculus of \cite{Melrose} enables us to make this kind of
argument precise.  Thus in \S\ref{conormal} we show that if $U= 1+ u$,
then $u$ is {\em conormal} at $\rinv=0$, which is to say that $u$ and
all derivatives of the form
$(\rinv\p_\rinv)^m \nabla_\omega^n u $ are continuous as  $\rinv \to
0.$
(This condition is strictly weaker than $u$ being smooth near $\rinv=0$.)
In \S\ref{unique},  we show that it is not possible for $U$ to
approach $1$ faster than $r^{-\mu}$ for {\em every} $\mu > 0$ unless $U=1$ in
$\RR^3$.
Finally in \S\ref{refined} we combine this fact with another
application of the $b$-calculus to show the existence of a non-trivial
asymptotic
expansion in powers of $r^{-1}$ and $\log r$ (or equivalently in
powers of $\rinv$ and $\log\rinv$).

\subsection{Rewriting the Skyrme Equation}
\label{rewrite}
To begin with we work near a fixed point of $\RR^3$, which we may as
well take to be the origin $0$. By replacing $U(x)$ by $U(0)^{-1}U(x)$
we can assume that $U(0)=1$. Write
\begin{equation}\label{1.17.10}
U(x) = 1 +u(x),
\end{equation}
so that $u(0)=0$.
Because $U$ is unitary, the $2\times 2$ complex matrix $u$ will
satisfy the algebraic constraints
\begin{equation} \label{2.17.10}
u + u^\dag + uu^\dag = 0,\;\; \tr(u) + \det u = 0.
\end{equation}
In particular $u$ is neither exactly skew-hermitian nor trace-free.
Then
\begin{equation}\label{1.9.10}
L_i = \p_i u + u^\dag \p_i u
\end{equation}
and
\begin{equation}\label{2.9.10}
\p_iL_i = (1+u^\dag)\Delta u + \p_i u^\dag \p_i u.
\end{equation}
The cubic term in the currents can be written
\begin{equation}\label{3.9.10}
\frac{1}{2}L_j[L_j,L_i]
= \frac{1}{2}(\p_j u + u^\dag \p_j u)
[\p_j u + u^\dag \p_j u,
\p_i u + u^\dag \p_i u].
\end{equation}
Taking the divergence, and using the notation
\begin{equation}\label{4.9.10}
v_{ij} = \p_i u^\dag \p_j u,
\end{equation}
we obtain
\begin{equation}\label{5.9.10}
\frac{1}{2}L_j[L_j,L_i] = (1+u^\dag)(T+F),
\end{equation}
where
\begin{equation}\label{6.9.10}
T=T(u,\p u, \p^2 u )
=
\frac{1+u}{2}(L_j[L_j,(1+u^\dag)\Delta u] + L_j[(1+u^\dag)\p_i u\p_j u,L_i])
\end{equation}
and
\begin{equation}\label{7.9.10}
F=F(u, \p u)=\frac{1+u}{2}(v_{ij}[L_i,L_j] +
L_j[L_j,v_{ii}] + L_j[(1+u^\dag)v_{ij},L_i]).
\end{equation}
The nonlinear terms have been divided here so that $T$ is
polynomial in the first derivatives of $u$ and linear in its second
derivatives, while $F$ only contains $u$ and its first derivatives.

With this notation,  the full Skyrme equation can be written
\begin{equation}\label{8.9.10}
P(u,\p u, \p^2 u) = Q(u,\p u) + F(u,\p u)
\end{equation}
where
\begin{equation}\label{9.9.10}
P(u,\p u,\p^2 u) = \Delta u + T(u,\p u, \p^2u)
\mbox{ and }Q(u,\p u) =  -(1+u)v_{ii}.
\end{equation}
Notice that $Q$ is (approximately) quadratic in $u$,  $T$ is
of degree three and $F$ is of degree four.
Note also that $T$ is linear in $\p_i\p_j u$ (and quadratic in
$\p_i u$).

Because of the assumed H\"older continuity of the currents, the
coefficients of the differential operator
\begin{equation}\label{3.17.10}
f\longmapsto P(u,\p u,\p^2 f)
\end{equation}
are H\"older continuous, and this operator is linear and elliptic in a 
small  neighbourhood $K$ of $0$.
 Moreover, the  RHS $Q+F$ of (\ref{8.9.10}) is also in
 $C^{0,\alpha}_b(K)$, so
by elliptic regularity, it follows that $u\in C^{2,\alpha}_b(K)$. Then the
currents are in  $C^{1,\alpha}_b(K)$ and the process continues to show that $u$
is in $C^{k,\alpha}_b(K)$ for every  $k$. Thus $u$ is smooth near $0$.
Since the point $0$ was arbitrary, the argument shows that $u$ is
smooth in $\RR^3$.

\subsection{Boundary regularity}

In order to analyze the asymptotic behaviour of the field $U$, we
shall make a transformation of the problem which involves passing from
$\RR^3$ to a `compactification' $\oR$
in which the sphere at infinity becomes a genuine boundary. This is
easily achieved by introducing the coordinate $\rinv=1/r$ along with
angular coordinates $\theta$ and $\varphi$ in $\RR^3$. Then in the
space $[0,\infty)\times S^2$ with coordinates $(\rinv,\theta,\varphi)$, the
boundary $\rinv=0$ corresponds to $r=\infty$ and $\theta$ and $\varphi$ give
coordinates on the boundary, which is the 2-sphere `at infinity'.
$\oR$ is defined to be the union of $\RR^3$ with the $2$-sphere at
infinity attached in this way.

 It
can be cumbersome to work with explicit coordinates on the $2$-sphere,
so we again use $\omega$ for points on $S^2$ and represent
any point other than the
origin of $\oR$   in the form
$(\rinv,\omega)$. 

Next, introduce rescaled derivatives
\begin{equation}\label{10.9.10}
D_i = r\p_i = \frac{1}{\rinv}\p_i.
\end{equation}
These vector fields have the property that they are linear
combinations, with coefficients that are smooth, down to
$\rinv=0$, of the  basic vector fields $\rinv\p_\rinv$, $\p_\theta$ and
$\p_\varphi$.
The Euclidean Laplacian takes the form
\begin{equation}\label{11.9.10}
\Delta  = \rinv^2\Delta_b
\end{equation}
with  $\Delta_b$ defined in (\ref{2.1.11}).

Now we write $U = 1 +u$ for large $r$ (that is, for small positive
$\rinv$) and make the rescalings of (\ref{8.9.10})
suggested by (\ref{10.9.10}) and
(\ref{11.9.10}). The result is a `$b$' version of the Skyrme equation,
\begin{equation}\label{5.17.10}
P_b(u,D  u, D^2 u) = Q_b(u,D u) + \rinv^2 F_b(u,D u),
\end{equation}
where
\begin{eqnarray}
P_b(u,D u,D^2 u) &=& \Delta_b u + \rinv^2 T_b(u,D u,D^2u),
\label{5b.17.10} \\
Q_b(u,D u) &=&  -(1+u)D_i u^\dag D_i u,
\end{eqnarray}
these terms being got by replacing $\p_i$ by the rescaled derivative
$D_i$ wherever they occur.

The reason for reformulating the equation in this way is that there
is a well-established theory, called the $b$-calculus,
which can be used to analyze equations of this kind  \cite{Melrose}.
 The $b$-calculus is concerned with $b$-differential
operators. In the present situation, a
$b$-differential operator is just a differential operator of the form
\begin{equation} \label{2.10.10}
P=\sum_{a+b+c\leq m}
C_{abc}(\rinv,\omega)(\rinv\p_\rinv)^a\p_\theta^b\p_\varphi^c,
\end{equation}
where the coefficients $C_{abc}$ are smooth up to the boundary
$\rinv=0$. From (\ref{2.1.11}) 
it is clear that the rescaled Laplacian $\Delta_b$ is an example of
such an operator. The set of all such operators will be denoted by
$\Diff_b$, those of order $k$ by $\Diff_b^k$.

The first aspect of this theory that is needed is the
counterpart of the elliptic regularity for H\"older spaces that we used
in the previous subsection. For this we need $b$-H\"older spaces, 
 already introduced in (\ref{10.1.11}).  With the new
variable $\rinv$, we have
$f(\rinv,\omega)\in C_b^{0,\alpha}$ if
\begin{equation} \label{1.10.10}
\sup_{x\in \RR^3} |f| + \sup_{(\rinv,\omega)\not= (\rinv',\omega')}
\frac{ (\rinv+ \rinv')^\alpha|f(\rinv,\omega) -
f(\rinv',\omega')|}{
|\rinv-\rinv'|^\alpha + (\rinv+ \rinv')^\alpha
d(\omega,\omega')^\alpha}
<\infty.
\end{equation}
Now put
$$
C^{k,\alpha}_b = \{f: Lf \in C^{0,\alpha}_b\mbox{ for all
$b$-differential operators }L \in \Diff_b^k\}
$$
and
$$
\cA = \{u: Lu\in C_b^{0,\alpha}\mbox{ for all }
L\in \Diff_b\}.
$$
In order to force functions to decay as $\rinv\to 0$, we introduce
{\em weighted} versions of these spaces,
$$
\rinv^\delta C_b^{k,\alpha} = \{u = \rinv^\delta v: v\in
C_b^{k,\alpha}\},\;\;
\rinv^\delta \cA = \{u = \rinv^\delta v: v\in \cA\}.
$$
The following is a very special case of elliptic regularity for
$b$-differential operators \cite{Mazzeo}.
\begin{theorem}
Consider the differential operator $P = \Delta_b + \rinv^2 E$,
where $E$ is a second-order differential operator with coefficients
smooth up to the boundary $\rinv=0$. Suppose that $Pu = f$ near
$\rinv=0$, with $u$ and $f$ in $\rinv^{\delta}C_b^{0,\alpha}$. Then
if $\delta$ is not an integer, it follows that $u\in \rinv^\delta
C_b^{2,\alpha}$.
\end{theorem}

We want to apply this to the Skyrme equation, written in the form
(\ref{5b.17.10}). After the last section, we know that the
coefficients of the perturbing term $E$ are smooth for $\rinv>0$,
but all we know at $\rinv=0$ is the original assumption that the
currents are in $\rinv^\delta C_b^{0,\alpha}$. The elliptic regularity
statement still holds in this case (though this statement does not
seem to be available in the literature). Applying this result, for
$\delta$ a small positive number, we obtain first that $u\in
\rinv^\delta C_b^{2,\alpha}$, which gives that the coefficients are in
$\rinv^{\delta}C_b^{1,\alpha}$, so $u\in \rinv^\delta C_b^{3,\alpha}$
and so on. Iterating, we find $u\in \rinv^{\delta}\cA$.

This is a major step forward: in particular, the angular variation of
$u$ is now very well controlled. However $u$ may still be very far
from being smooth up to the boundary or having an
asymptotic expansion there. Indeed, horrors like $\rinv^\delta \sin \log
\rinv$ lie in $\rinv^\delta \cA$.
\label{conormal}
\subsection{Power-law decay of $u$}
\label{unique}
In this section and the next we establish that a topologically 
non-trivial  solution of the Skyrme equation 
 must have a non-trivial asymptotic
expansion in powers of $1/r$.  We show first that
it is not possible for a
topologically non-trivial solution  to approach $1$ faster than every
power of $1/r$. This result is then fed into an iterative analysis of
the equation in the next section. These two sections can, however, be read
in  either order. The main result of this section is as follows.

\begin{theorem} If the topological charge of the Skyrme field $U$ is
non-zero and if it satisfies the hypotheses of Theorem~\ref{asymps}, then
there exists some $\mu \in \RR^+$ such that $r^\mu u(r,\omega)$ 
does {\em not} tend
to zero as $r\rightarrow \infty$.
\label{nt}
\end{theorem}

To set this result in a more general context, recall that
a partial differential equation $Lu = 0$ is said to have the unique
continuation property at a point $0$, say, if the following is true:
\be \label{5.1.11}
\mbox{If all derivatives of $u$ vanish at $0$, then }u=0\mbox{ in some
neighbourhood of $0$.}
\ee
The methods used to establish that a second-order PDE has the unique
continuation property establish analogous statements with somewhat weaker
hypotheses. For example, a simplified version of Theorem~17.2.6 of
\cite{Hormander} is as follows.

\begin{theorem} Let $a_{jk}(x)$ be smooth and positive-definite in a
neighbourhood $X$ of $\,\,0$, and suppose that $a_{jk}(0)=
\delta_{jk}$. Suppose that for $x\in X$, the smooth function $u$ satisfies
\be\label{6.1.11}
|a_{jk}(x)\p_j\p_k u| \leq A \left(|u(x)| + |\nabla u(x)|\right)
\ee
for some constant $A$, and
\be \label{7.1.11}
|x|^{-\mu} |u(x)|\to 0 \mbox{ as} \quad 
|x|\to 0\quad  \mbox{for {\em every} } \mu
\in \RR^+.
\ee
Then $u=0$ identically in a neighbourhood of $0$.
\label{8.1.11}\end{theorem}

Using this result we can prove the following:

\begin{theorem} Let $U$ satisfy the hypotheses of
Theorem~(\ref{asymps}). Then if the topological charge of $U$ is
non-zero, $U$ cannot be constant in any open subset of $\RR^3$.
\end{theorem}

\noindent{\bf Proof} \,\,\, Define
\begin{equation}
W = \{x\in \RR^3: U(y)= U(x) \mbox{ for all }y\mbox{ in some
neighbourhood of }x\}.
\end{equation}
Then $W$ is  open by definition.  By Theorem~\ref{8.1.11}, $W$ is also
closed. To see this, suppose that $x_n\in W$, $x_n\to x_0\in
\RR^3$. Then all derivatives of $U$ are zero at $x_0$, so
(\ref{7.1.11}) holds (with $0$ replaced by $x_0$).  The Skyrme
equation, in the form (\ref{8.9.10}), implies a differential inequality
of the form (\ref{6.1.11}) in a neighbourhood of $x_0$.  
It follows\footnote{Unique
continuation theorems do not extend wholesale to systems. However, in
our case, the leading term is the Laplacian and the other second-order
terms $C(u,\p u, \p^2 u)$ are non-scalar but small near $x_0$. The
proof of Theorem~\ref{8.1.11} goes through in this case.} that $U$ is
identically constant in a neighbourhood of $x_0$, so that $x_0 \in
W$.  Since $\RR^3$ is connected, $W= \emptyset$ or $W= \RR^3$.\hfill\halmos

We will use this theorem to give an indirect proof of Theorem~\ref{nt}.
Suppose $r^\mu u(x)  \to 0$  as $r\to \infty$   for every $\mu $.
Adapating Theorem~\ref{8.1.11} we shall show  that then $u=0$ 
for {\em all} sufficiently large $r$. Thus $U$ is constant
in an open set, hence by the previous result constant everywhere.

So consider the $b$-differential operator
\be\label{1.3.11}
P_b = \Delta_b + \rinv^2 T_b
\ee
on $\oR$, where the coefficients of $T_b$ are in $C_b^{0,\alpha}$.  By
rescaling   Theorem~17.2.6 in \cite{Hormander}, we obtain
\begin{theorem}
Suppose that
\be \label{2.3.11}
|P_bu (\rinv,\omega)| \leq A\rinv^\delta
(|u(\rinv,\omega)| + |D u(\rinv,\omega)|)
\mbox{ for all }0<\rinv < \rinv_0,\omega\in S^2
\ee
and that
\be\label{3.3.11}
\rinv^{-\mu}|u(\rinv,\omega)| \to 0 
\mbox{ as }\rinv\to 0\mbox{ for all } \mu.
\ee
Then $u(\rinv,\omega)=0$ for $0\leq \rinv < \rinv_1$, where $\rinv_1$
is some small positive number.
\label{buniq}\end{theorem}

The `$b$' version (\ref{5.17.10}) of the Skyrme equation implies a
differential inequality of the
form (\ref{2.3.11}), just as before. It follows that if $u$ decays
faster than any power of $r$, then $u=0$ for all sufficiently large
$r$.
By the remarks before the statement of Theorem~\ref{buniq}, the proof
of Theorem~\ref{nt} is now complete.

\subsection{Refined regularity, asymptotic expansions}\label{refined}

The main result of this section can be stated as follows:
\begin{theorem}\label{fthm}
Let $U=1+u$ satisfy the hypotheses of Theorem~\ref{asymps}.  Then
there is some
integer $M\geq 1$ and an asymptotic expansion
\be \label{4.3.11}
u \sim \sum_{j=M}^{2M}Y_{j}(\omega)r^{-(j+1)}+
\sum_{j=2M+1}^\infty r^{-(j+1)}w_j(\omega,\log r)\mbox{ for large }r.
\ee
Here the $Y_j$ are Lie-algebra valued spherical harmonics,
\be\label{8.3.11}
\Delta_\omega Y_j = -j(j+1) Y_j,
\ee
and $Y_M\not=0$,
so that the piece $\sum_{j=M}^{2M}Y_{j}(\omega)
r^{-(j+1)}$ is a non-zero harmonic
function. The
functions $w_j$ are smooth in $\omega$ and polynomial in $\log r$.
\end{theorem}

It will follow from the proof that the terms $w_{2M+1}$ to $w_{3M+1}$
are of at most degree $1$ in $\log r$, the terms $w_{3M+2}$ to
$w_{4M+2}$ are of at most degree $2$ in $\log r$ and so on. We remark
also that the asymptotic expansion can safely be differentiated term
by term to give asymptotic expansions of all derivatives of $u$.

In order to motivate the proof, consider
the equation
\be \label{5.3.11}
\Delta_b u = f
\ee
where $u$ and $f$ are defined for small   $\rinv$. If $f$ has
the form $f = \rinv^{\lambda}g(\omega)$, then a solution can be found
as follows. Expand $g$ as a sum of spherical harmonics,
$$
g = \sum_{j=0}^\infty g_j,\mbox{ where }\Delta_\omega g_j = -j(j+1)g_j
$$
and seek a solution $u = \sum u_j(s)g_j$. Then $u_j$ must satisfy
$$
[(\rinv\p_\rinv)^2 - (\rinv\p_\rinv) -j(j+1)]u_j(\rinv) = \rinv^\lambda.
$$
This is solved by
$$
u_j(\rinv) = \frac{\rinv^{\lambda}}{\lambda(\lambda-1) -j(j+1)}
$$
provided there is no resonance, that is to say
$$
\lambda \not= -j, j+1.
$$
In the resonant case, $u_j(\rinv)$ has the form
$\rinv^\lambda(A +  B\log\rinv)$. The general solution is
obtained by combining this with an arbitrary solution of the
homogeneous equation $\Delta_b v =0$. If we require $u\to 0$ as
$\rinv\to 0$, then $v$ must itself go to zero and hence will be a sum
of multipoles, $v = \sum_{j=0}^\infty Y_{j}(\omega)\rinv^{j+1}$, where
$Y_j$ satisfies (\ref{8.3.11}).

The $b$-calculus extends results of this kind to functions in
$\rinv^\delta \cA$ (which, as we have seen, can be far from having
expansions in powers of $\rinv$). In order to summarize the needed
results, write
\be \label{6.3.11}
f = \cO(\rinv^{\delta})\mbox{ instead of }f\in \rinv^\delta \cA.
\ee
This conforms to the use of the `$\cO$'-notation in the rest of the
paper, but has the additional property
\be \label{7.3.11}
\mbox{If }f = \cO(\rinv^\delta)\mbox{ then }Lf = \cO(\rinv^\delta)\mbox{ for
all }L\in \Diff_b.
\ee

\begin{lemma} Suppose that $u$ and $f$ defined near $\rinv=0$ satisfy
(\ref{5.3.11}). If
$u = \cO(\rinv^\delta)$ and $f =
\cO(\rinv^{n+\delta})$, where $\delta >0$ and $n$ is a positive
integer, then
\be \label{11b.17.10}
u = \sum_{j=0}^{n-1} Y_{j}(\omega)\rinv^{j+1} +  \cO(\rinv^{n+\delta})
\ee
where $Y_j$ satisfies (\ref{8.3.11}). In particular the sum on the RHS
is harmonic
\begin{equation}\label{12.17.10}
\Delta_b \sum_{j=0}^{n-1} Y_{j}(\omega)\rinv^{j+1} = 0.
\end{equation}
\label{l1}\end{lemma}

\begin{lemma}\label{l2}
Suppose that $u$ and $f$ defined near $\rinv=0$ satisfy
(\ref{5.3.11}). If
$u = \cO(\rinv^\delta)$ and
\begin{equation} \label{1.19.10}
f =  \sum_{j=0}^{n-1} f_j(\omega,\log \rinv) \rinv^{j+1}+
\cO(\rinv^{n+\delta}),
\end{equation}
where $\delta >0$, $n$ is  a positive integer and $f_j$ is polynomial
of degree $m_j$ in $\log s$, then
\begin{equation}\label{2.19.10}
u = \sum _{j=0}^{n-1} w_j(\omega,\log \rinv)\rinv^{j+1} +
\cO(\rinv^{n+\delta})
\end{equation}
where $w_j$ is a polynomial of degree $m_j+1$ in $\log \rinv$.
\end{lemma}

These results will be applied to the Skyrme equation, now rewritten as
\begin{equation}\label{13.17.10}
\Delta_b  u  =  Z_b(u): =Q_b(u) - \rinv^2 T_b(u) + \rinv^2F_b(u).
\end{equation}

\begin{lemma} Suppose that $U=1+u$ satisfies the Skyrme equation and
$u = \cO(\rinv^\delta)$. Then $u = \cO(\rinv^{2})$.
\end{lemma}

\noindent{\bf Proof:} On the RHS of (\ref{13.17.10}),
$Q_b$ is quadratic in $D u$
and the other terms
in $Z_b$ are of even higher degree. Hence $Z_b = \cO(\rinv^{2\delta})$.
Applying Lemma~\ref{l1} we obtain that $u = \cO(\rinv^{2\delta}) +
\cO(\rinv)$. If $2\delta<1$, we can iterate this argument to obtain
eventually $u\in \cO(\rinv)$. In \cite{Mantonn}, it is shown that a
solution of the Skyrme equation 
 cannot have leading term $1/r = \rinv$ in its asymptotic
expansion. It follows that $u = \cO(\rinv^{1+\delta})$ (for a possibly
smaller $\delta>0$). Hence $Z_b(u) = \cO(\rinv^{2+2\delta})$ and so,
applying Lemma~\ref{l1} again, $u = Y_1 \rinv^2 +
\cO(\rinv^{2+\delta})$.\hfill\halmos

Combining Theorem~\ref{nt} with Lemma~\ref{l1}, we see that there
exists an integer $M\geq 1$ with the property that
\be\label{9.3.11}
u  = Y_M(\omega) \rinv^{M+1} +\cO(\rinv^{M+1+ \delta}),
\ee
where $Y_{M}$ is a {\em non-vanishing} spherical harmonic.  We can now
complete the proof of Theorem~\ref{fthm} in the following iterative
fashion.  From the structure of $Z_b(u)$ it follows from
(\ref{9.3.11}) that
\be\label{9a.3.11}
Z_b(u) = f_{2M+1}\rinv^{2M+2}+\cO(\rinv^{2M+2+\delta}).
\ee
Applying Lemma~\ref{l2},
\be\label{10.3.11}
u = \sum_{j=M}^{2M}Y_j(\omega)
\rinv^{j+1} + w_{2M+1}(\omega,\log \rinv)\rinv^{2M+2}
+  \cO(\rinv^{2M+2+\delta}),
\ee
where $w_{2M+1}(\omega, \log \rinv)$ 
is of degree at most $1$ in $\log \rinv$. Now this
expression for $u$ is substituted into $Z_b$, to give
\be \label{4.19.10}
Z_b = \sum_{2M+1}^{3M+1}f_j(\omega)\rinv^{j+1} + f_{3M+2}(\omega,\log
\rinv)\rinv^{3M+3} +  +\cO(\rinv^{3M + 3 +\delta})
\ee
where  $f_{3M+2}$ is of degree at most $1$ in $\log
\rinv$. 
Now apply Lemma~\ref{l2} to get
\bea
u=\sum_{j=M}^{2M}Y_j(\omega)\rinv^{j+1} + \sum_{j=2M+1}^{3M+1}w_j(\omega,\log
\rinv) s^{j+1} + w_{3M+2}(\omega,\log s) s^{3M+3}
+\cO(\rinv^{3M+3+\delta}),
\eea
where the functions $w_{2M+1},...,w_{3M+1}$ are of degree at most 1 
in $\log \rinv$ and $w_{3M+2}$ is of degree at most 2 in $\log \rinv$.
Carrying on in this way we obtain a complete asymptotic expansion.

To complete the proof of 
Theorem~\ref{fthm} note finally 
that the first $M+1$ terms in the expansion are actually
the pion field. Indeed, if the expansion (\ref{4.3.11}) is
substituted into the algebraic constraints (\ref{2.17.10}) we see
that the harmonic piece $\sum_M^{2M} Y_j\rinv^{j+1}$ is skew-adjoint
and trace-free---the quadratic corrections enter at order $\rinv^{2M+2}$.
\hfill \halmos

The upshot of this section is that every  Skyrme
soliton has a leading
Lie-algebra valued multipole field (called a $2^M$-pole)
\bea
\label{genmultipole}
u_M(x) =i\tau_a\sum_{m=-M}^M\frac{4\pi}{2M+1} Q^a_{Mm}\frac{Y_{Mm}
(\theta,\varphi)}{r^{M+1}},
\eea
where $Y_{Mm}$ are the usual spherical harmonics on $S^2$, see
appendix A. The leading
multipole  moment $Q^a_{Mm}$ is independent of the location of
the Skyrme soliton,  and is acted on naturally by rotations
and iso-rotations. It is a key ingredient in  the calculations of  the
following sections. As already mentioned
one can show that
Skyrme solitons never have asymptotic monopole fields \cite{Mantonn}.
The  leading multipole  field of  the  $B=1$ hedgehog (\ref{hedge})
is a triplet of dipoles,  and dipoles are known to occur as leading
multipoles in a number of Skyrme solitons.
The highest leading multipole known from numerical work is an
octupole, which occurs in a  $B=7$ configuration with icosahedral
symmetry \cite{BS1}.

\section{The interaction energy of two Skyrme solitons}

Suppose we have Skyrme solitons  $U^{(1)}$ and
$U^{(2)}$  of  degrees  $k$
and $l$ which minimise the energy in the sectors ${\cal C}_k$ and
${\cal C}_l$.
Since the total energies  of
$U^{(1)}$ and $U^{(2)}$ are finite there must
be balls  $B_1$ and $B_2$ in $\RR^3$ so that most of the energy
 of $U^{(1)}$ and $U^{(2)}$  is concentrated in,
respectively, $B_1$ and $B_2$. Outside the balls $B_1$ and $B_2$
the asymptotic analysis of the previous section applies. Suppose
that the leading multipole of $U^{(1)}$ is a $2^M$-pole and the
leading multipole of $U^{(2)}$ is a $2^N$-pole. Denoting the
radii of $B_1$ and $B_2$ by $D_1$ and $D_2$, and with
the abbreviation (\ref{genmultipole}) for a generic
Lie-algebra valued multipole field we have
\bea
U^{(1)}(x) \sim 1+u_M(x)\quad\mbox{for}\quad  x\not \in B_1
\eea
and
\bea
U^{(2)}(x) \sim 1+v_N(x)\quad\mbox{for}\quad x\not \in B_2.
\eea

Using
the translational invariance of the Skyrme energy functional we can
assume  without loss of generality that $B_1$  is centred at
$X_+=(0,0,R/2)$  and   that $B_2$ is centred at $X_-=(0,0,-R/2)$, where $R$ is
so large that $B_1$ and $B_2$  do not overlap, i.e. $R>D_1+D_2$.
The  parameter $R$  will be interpreted  as the separation of the Skyrme
solitons. There is clearly an ambiguity in the definition of such
a separation parameter, but this does not affect our calculation
of leading terms in the limit where $R$ becomes large.
Then we define the following product configuration
\bea
\label{product}
U_R(x)=U^{(1)}(x)U^{(2)}(x).
\eea
This configuration  has degree $k+l$ and we shall see shortly that
its energy is finite, so that  $U_R \in {\cal C}_{k+l}$.

Our goal is to study  the energy of the product configuration $U_R$
perturbatively in the limit of large $R$
 and to compute the leading terms in powers of $1/R$.
A similar calculation  for  moving and spinning  Skyrmions
was performed in  \cite{Schroers}, where some further details
are given.
Let $\Li$ and $\LLbar$ be the currents (\ref{Lc}) and (\ref{LLc})
constructed out of $U^{(1)}$, and $\R$ and $\Rbar$ be the currents
(\ref{Rc}) and (\ref{RRc})
constructed out of $U^{(2)}$. Then    one computes
\bea
\label{inen}
E[U_R]&=&E[U^{(1)}] +E[U^{(2)}] + W_{2}+ W_4 .
\eea
The energies $ E[U^{(1)}]$ and $ E[U^{(2)}]$ are simply the
energies of the Skyrme solitons $U^{(1)}$ and $U^{(2)}$ and therefore
independent of $R$.
The  interaction terms  $W_{2}$  and
$W_4$  are  given by integrals  over $\RR^3$
\bea
W_{2}=\int d^3x \,w_2 \quad \mbox{and} \quad W_4=\int d^3x \,w_4,
\eea
with integrands
\bea
w_2 = \,\tr\left( \Li\Rbar+\LLbar \R-\Li\R\right)
\eea
and
\bea
 w_4= -\frac{1}{8} \tr\left(
 [\Li,\Rj][\Li,\Rj]+ [\Li,\Rj][\R,\Lj]) + [\Li,\Lj][\R,\Rj]\right).
\eea

We shall see shortly that the term $W_2$   contains
 the leading contribution to the interaction energy.
However, for the presentation of
our method of computation it is more convenient to begin with
the quartic term  $W_4$.
We split the integration region $\RR^3$ into the balls $B_1$
and $B_2$ and the  complement $C=\RR^3 -(B_1\cup B_2)$.
To illustrate our method, consider the integral
\bea
I  = -\frac{1}{8} \int_{B_1} d^3x\,\tr\left(
 [\Li,\Rj][\Li,\Rj]\right) \leq  \frac{1}{4} \int_{B_1} d^3x\,
\tr\left(\Li\Li\right)\tr\left(\Rj\Rj\right).
\eea
The currents $\Li$  are smooth functions and hence bounded
on the compact domain $B_1$. Therefore
\bea
\label{remaining}
|I|\leq  -  K \int _{B_1} d^3x\, \tr\left(\Rj\Rj\right).
\eea
for some positive constant $K$. Since   $B_1$ is far away from the centre
of soliton $U^{(2)}$, the leading contribution to the integral
(\ref{remaining})
is obtained  by  replacing   $\Rj$ by
the leading multipole  component  $-i\partial_j v_N$.
Using
\bea
|\partial_j v_N|(x) \leq \frac{K'}{|x-X_-|^{N+2}}
\eea
for a further positive constant $K'$
we conclude that
\bea
I={\cal O}\left(\frac{1}{R^{2N+4}}\right).
\eea
A similar calculation for the other terms in $W_4$ shows that 
\bea
\label{bone}
\int_{B_1} d^3x\,w_4 =
{\cal O}\left(\frac{1}{R^{2N+4}}\right).
\eea
Considering the contribution from $B_2$ we find by the same argument
\bea
\label{btwo}
\int_{B_2} d^3x\,w_4 =
{\cal O}\left(\frac{1}{R^{2M+4}}\right).
\eea
Finally the remaining integral over $C$ can be estimated as follows.
Define
\bea
F(R)= \int_C d^3x \, \frac{1}{|x-X_+|^{2M+4}}
\frac{1}{|x-X_-|^{2N+4}},
\eea
noting that the integral converges for all values of $R>D_1+D_2$.
Then there is a positive constant $K''$ such that
\bea
\int_C d^3x \,w_4 < K''\, F(R).
\eea
The large $R$ behaviour of $F$ can be estimated by a scaling
analysis. Fix $R_0>D_1+D_2$ and consider $R>R_0$. Changing integration
variables $x\mapsto (R_0/R)x$ one computes
\bea
F(R)=\left(\frac{R_0}{R}\right)^{2M+2N+5}\left[
F(R_0)+
 \int_{S_1(R)\cup S_2(R)} d^3x \, \frac{1}{|x-X_+|^{2M+4}}
\frac{1}{|x-X_-|^{2N+4}}\right],
\eea
where $S_1(R)$  and $S_2(R)$ are the thick shells
\bea
S_1(R)&=&\{x\in \RR^3|(R_0/R)D_1\leq |(x_1,x_2,x_3-(R_0/2)|<D_1
\}\nonumber \\
S_2(R)&=&\{x\in \RR^3|(R_0/R)D_2\leq
 |(x_1,x_2,x_3+(R_0/2)|<D_2
\}
\eea
which converge to punctured balls
\bea
B^0_1&=&\{x\in \RR^3-\{(0,0,R_0/2)\}|\,|(x_1,x_2,x_3-(R_0/2)|<D_1
\}\nonumber \\
B^0_2&=&\{x\in \RR^3-\{(0,0,-R_0/2)\}|
 |(x_1,x_2,x_3+(R_0/2)|<D_2
\}
\eea
in the limit $R\rightarrow \infty$. In that limit the integral over
$S_1(R)$ diverges like $R^{2N+1}$ and that over $S_2(R)$ like
$R^{2M+1}$. Combining this with with the factor $R^{-(2M+2N+5)}$
we deduce that $F(R)$ decays for large $R$ like $ R^{-(2N+4)}$
and  $R^{-(2M+4)}$, just like the contributions (\ref{bone})
and (\ref{btwo}). Combining all the terms,
we conclude that the leading terms in $W_4$  decay for large $R$
according to $R^{-(2N+4)}$ and  $R^{-(2M+4)}$.

In order to evaluate $W_2$
 we first divide the region of integration into the half-spaces
\bea
H^+=\{x \in \RR^3\, | \,x_3 > 0\}\qquad \mbox{and }\qquad
H^-=\{x\in \RR^3\,|\,x_3 <0 \}.
\eea
 In $H^+$ we replace  $U^{(2)}$ by the leading term $1+v_N$
and in $H^-$ we replace $U^{(1)}$
 by the leading contribution  $1+u_M$. The result
is
\bea
\label{wtwo}
W_{2}&\approx& \int_{H^+} d^3x
\, \frac 1 4 \tr\left( \Li [\partial_j v_N,[\partial_j v_N,\partial_i
  v_N]] \right)
-\tr\left(\partial_i v_N\LLbar\right) \nonumber \\
&-& \int_{H^-}d^3x \,
 \frac 1 4 \tr\left( \R [\partial_j u_M,[\partial_j u_M,\partial_i
   u_M]] \right)
+ \tr\left( \partial_i u_M\Rbar\right).
\eea
Now integrating by parts and using the Euler-Lagrange equations
$\partial_i\LLbar=\partial_i\Rbar=0$
for the individual Skyrme solitons
we convert two of the terms into an area
integral
\bea
 -\int_{H^+} d^3x \, \tr\left(\partial_i v_N\LLbar\right)
+ \int_{H^-}d^3x\!\!\!\!\!\!\!\!\!\!
&& \!\!\!\tr\left( \partial_i u_M\Rbar\right)
\nonumber \\
&=&\int_{x_3=0}dx_1 dx_2 \,\tr\left(v_N \tilde L_3^{(1)}
 +u_M\tilde R_3^{(2)}\right).
\eea
Since the $x_1x_2$-plane is far away
from both Skyrme solitons the  leading contribution to this area
integral can be expressed
entirely in  terms of the asymptotic fields:
\bea
\label{key}
\Delta E  = 2\sum_{a=1}^3 \int_{x_3=0}dx_1 dx_2 \,
(u_M^a\partial_3v_N^a-v_N^a\partial_3u_M^a).
\eea
A simple scaling analysis shows that  $\Delta E$ falls off like
$R^{-(N+M+1)} $ for large $R$. We skip the details here
because we shall show
how to evaluate $\Delta E$ {\it exactly} in the next section.
The remaining terms in (\ref{wtwo}) can be estimated with the
techniques used in estimating $W_2$. The result is
\bea
W_2 =\Delta E + {\cal O}\left(\frac {1} {R^{3N+6}}\right) +
{\cal O}\left(\frac {1} {R^{3M+6}}\right)
\eea
Combining all terms in (\ref{inen}) we conclude that
\bea
\label{inenn}
E[U_R]=E[U^{(1)}] +E[U^{(2)}] + \Delta E + {\cal O}\left(\frac {1}
  {R^{2N+4}}\right) +
{\cal O}\left(\frac {1} {R^{2M+4}}\right).
\eea
Note that $\Delta E$ is the leading contribution to the interaction
energy if $|N-M|\leq 2$ i.e. if the   orders  of the leading
multipoles  of the two Skyrme solitons
differ by at most two. We will comment on the validity of
this assumption at the end of this paper.

\section{Harmonic functions and their interaction energy}

In order to compute the interaction energy $\Delta E$ we need to
derive some  general results about harmonic functions.
We define   the regions
\bea
H_\delta^-=\{x \in \RR^3\,|\,x_3 < \delta \}\qquad \mbox{and }\qquad
H_\delta^+=\{x\in \RR^3\,|\,x_3 > -\delta \},
\eea
where the positive parameter $\delta$ is introduced for technical reasons.
Then  we introduce the spaces
\bea
{\cal H}^-=\{f: H^-_\delta\rightarrow \RR\,|\,\Delta f=0,\quad
\lim_{r\rightarrow \infty}f(x)=0\}
\eea
and
\bea
{\cal H}^+=\{g: H^+_\delta \rightarrow \RR\, | \,\Delta g=0,\quad
\lim_{r\rightarrow \infty}g(x)=0\}.
\eea
Elements of ${\cal H}^-$ tend to zero
at the boundary ``at infinity'' of $H_\delta^-$,
elements of ${\cal H}^+$ tend to zero at the boundary ``at infinity'' of
$H_\delta^+$.
No additional restriction is placed on behaviour  at
the  boundaries
$x_3=\pm \delta$.

For the calculations in this section it is  convenient to  split
$\RR^3$ into $\RR^2\times \RR$ and denote vectors in $\RR^2$
by bold letters, e.g. $\bx =(x_1,x_2)$. We then  write $x=(\bx,x_3)$.
The most general element of ${\cal H}^-$ can be written as
\bea
\label{mexp}
f(x)=\int \frac{d^2\bk}{(2\pi)^2 2k} \,\,p(\bk) \, \exp(i\bk\cd\bx +kx_3) ,
\eea
where $k=\sqrt{\bk^2}$ and the volume element $ {d^2 \bk}
/({(2\pi)^2 2k})$
arises from the combination of $d^3k$ with the delta-function
$\delta(\bk^2-k^2)$ which ensures that $f$ satisfies the
Laplace equation. Since  $f$ is real the Fourier
transform $p$ satisfies
\bea
\label{preal}
\bar p(\bk)=p(-\bk).
\eea
Similarly, the most general element of ${\cal H}^+$ can be written as
\bea
\label{pexp}
g(x)=\int \frac{d^2\bl}{(2\pi)^2 2l} \,\,  q(\bl)\, \exp(i\bl\cd\bx -lx_3),
\eea
with $l=\sqrt{\bl^2}$ and
\bea
\label{qreal}
\bar q(\bk)=q(-\bk).
\eea

There is a natural pairing between elements of the
${\cal H}^-$ and those of $  {\cal H}^+$
\bea
\label{pairing}
\langle f,g\rangle  =
\int_{x_3=0} dx_1dx_2\,\,(  g \partial_3 f -f\partial_3 g)
\qquad  \mbox{for} \quad
f\in {\cal H}^-, g\in {\cal H}^+,
 \eea
Using the expansion (\ref{mexp}) and (\ref{pexp}) we find, in terms of
the Fourier modes,
\bea
\label{inprod}
\langle f,g\rangle =
\int \frac{d^2\bk}{(2\pi)^2 2k} \,\,p(\bk) q(-\bk).
\eea
This pairing is of interest to us since
the  interaction energy (\ref{key})
is proportional to the sum over the pairings $\langle u^a_M, v^a_N\rangle$.
Therefore, we
also refer to the expressions  (\ref{pairing}) and (\ref{inprod})
 as the  interaction  energy of the harmonic functions $f$ and $g$.

It is clear that the pairing (\ref{pairing})
may vanish for some  pairs of
harmonic functions $f\in{\cal  H}^-$, $g\in {\cal H}^+$.
This happens for example if the support of the Fourier
transform $p$ is complementary to that of the Fourier transform $q$.
However, we shall now show that
the interaction energy  of multipoles
can always be made non-zero by rotating one of the functions.

\subsection{Multipole fields}
For $f\in {\cal H}^-$ we have an alternative expansion in
spherical harmonics $Y_{Mm}$
\bea
f(x)=\sum_{M\geq 0}\sum_{m=-M}^{M}\frac{4\pi }{(2M+1)}
Q_{Mm}\frac{Y_{Mm}(\theta_+,\varphi_+)}{r_+^{M+1}},
\eea
where $r_+=|x-X_+|$ and $(\theta_+,\varphi_+)$ are spherical
coordinates centred at $X_+=(0,0,R/2)$. In this section we only
need  to assume $R>0$, but in our applications we will be
interested in the large $R$ limit.
The coefficients $Q_{Mm}$ are called the multipole moments of the function $f$.
Assume that $f$ has non-vanishing  multipole moments  and suppose $M$ is the
smallest integer such that $Q_{Mm}\neq 0$ for some $m=-M,...,M$.
The function
\bea
f_M(x)=\frac{4\pi }{(2M+1)}
\sum_{m=-M}^{M}Q_{Mm}\frac{Y_{Mm}(\theta_+,\varphi_+)}{r_+^{M+1}}.
\eea
is   a $2^M$-pole field  and $Q_{Mm}$ are
the leading multipole moments of $f$.

It is often convenient to
write  multipole fields  in terms of partial derivatives
of the Coulomb potential centred at $X_+$:
\bea
\label{coulo}
\phi_+(x) =\frac{1}{r_+}=\frac{1}{\sqrt{\rho^2 +
(x_3-R/2)^2}},
\eea
where $\rho^2=\bx^2$.
The function
$\partial_3^{m_3} \partial_2^{m_2} \partial_1^{m_1}\,\,\phi_+(x)$
is a  $2^M$-pole field  if $m_1+m_2+m_3=M$. However, not of all of
the fields obtained in this way are independent.
 We introduce the complex derivatives
\bea
\partial =\frac{1}{2}(\partial_1-i\partial_2)\quad \mbox{and}
\quad {\bar \partial} =\frac{1}{2}
(\partial_1+i\partial_2)
\eea
and note that
$\Delta =\partial_3^2+4\partial\bar \partial$. Then, since $\Delta\phi_+(x)=0$
 we have
\bea
 \partial\bar \partial\,\phi_+(x)= -\frac{1}{4}\partial_3^2 \,\phi_+(x).
\eea
Thus a basis for  $2^M$-pole fields  is given by
$\partial_3^{m_3}\partial^n{\bar \partial}^{\bar n}\phi_+(x)$ where
$M=m_3+n+\bar n$  and  either $n$ or $ \bar n$ can be taken to be zero.
In  appendix A we  derive the exact relation between
the  functions $\partial_3^{M+m} {
  \partial}^{-m} \phi_+(x) $, $-M\leq m< 0$,  and
$ \partial_3^{M-m} {\bar \partial}^m \phi_+(x) $, $0\leq m\leq M$,
 on the one hand   and
  the spherical harmonics
$Y_{Mm}$ centred at $X_+$ on the other. The result is that
we have the
alternative  expansion of an  $2^M$-pole field
\bea
\label{umultipole}
f_M(x) = \sum_{-M\leq m\leq 0}
A_{Mm} \,\,\partial_3^{M+m} { \partial}^{-m} \phi_+(x) +
\sum_{1\leq m\leq M}
A_{Mm} \,\, \partial_3^{M-m} {\bar \partial}^m \phi_+(x),
\eea
where the coefficients $A_{Mm}$, $m=-M,...,M$
are directly proportional  to the multipole moments $Q_{Mm}$.
It follows from the results in appendix A  that
\bea
\label{rescone}
 A_{Mm}=\sqrt{\frac{4\pi}{2M+1}}
\frac{(-1)^{M+m}2^{m}}{\sqrt{(M-m)!(M+m)!}}
Q_{Mm}
\eea
for $m\geq 0$ and
\bea
\label{resctwo}
 A_{Mm}=\sqrt{\frac{4\pi}{2M+1}}
\frac{(-1)^{M}2^{|m|}}{\sqrt{(M-m)!(M+m)!}}
Q_{Mm}
\eea
for $m<0$.
Note in particular that the reality of  $f_{M}$ is equivalent to
\bea
A_{M(-m)}=\bar A_{Mm}.
\eea

Multipole fields have  a remarkably simple Fourier
transform, which will be important for us. We use  the representation
\bea
\label{coulfour}
\phi_+(x)=\int \frac{d^2\bk} {2\pi k}\,\,\,
e^{-k|x_3-\frac{R}{2}|} \, \exp({i\bk \cdot\, \bx}),
\eea
which can be verified as follows.
Exploiting  the invariance of $\phi_+$ under rotations in the
$x_1x_2$-plane we may assume that $\bx =(\rho,0)$.
Using polar coordinates $(k,\psi)$ for $\bk$
we  first carry out  the  $dk$ integration and then
the angular integration:
\bea
\int_0^{2\pi} \frac{ d\psi } {2\pi}\int_0^\infty dk\,
e^{-k|x_3-\frac{R}{2}|} \, \exp({i\bk \cdot\, \bx})
&=&\int_0^{2\pi} \frac{d\psi}{2\pi} \frac{-1}{(i\rho\cos\psi -|x_3-R/2|)}\\
&=&\oint_{S^1}\frac{dw}{2\pi i}\frac{-2}{(i\rho w^2 -2|x_3-R/2|w +i\rho)},
\eea
where we  changed variables to $w=e^{i\psi}$ in the last line.
Expanding the integrand in partial fractions and using
the residue theorem then yields the expression (\ref{coulo}).
We compute the Fourier transform of the multipole field  (\ref{umultipole})
by differentiating (\ref{coulfour})
 under the integral sign. Note  that
\bea
\partial \,\exp({i\bk\cdot \,\bx})= \frac {i}{2}ke^{-i\psi}
\exp({i\bk\cdot \,\bx})
\quad\mbox{and}
\quad  {\bar \partial}\, \exp({i\bk\cdot \,\bx})= \frac {i}{2}ke^{i\psi}
\exp({i\bk\cdot \,\bx}),
\eea
so we find
\bea
\label{un}
f_M(x) =\frac{1}{2\pi}
\int d^2\bk \,\,e^{-k(\frac{R}{2}-x_3)}\,k^{M-1}\sum_{-M\leq m \leq M}
\left(\frac{i}{2}\right)^{|m|}A_{Mm}e^{im\psi}\, \exp({i\bk\cdot \,\bx}).
\eea
Here we have used that $x_3<\delta$ so that in particular $x_3 <\frac{R}{2}$.
Thus with the normalisation (\ref{mexp}) we arrive at  the following simple
expression for the  Fourier transform
\bea
\label{fouru}
p_M(\bk)=4\pi e^{-k\frac{R}{2}}k^{M}\sum_{-M\leq m \leq M}
\left(\frac{i}{2}\right)^{|m|}A_{Mm}e^{im\psi}.
\eea
Remarkably, this function factorises into a $k$-dependent part
and  the function
\bea
\Theta(\psi)=\sum_{-M\leq m \leq M}
\left(\frac{i}{2}\right)^{|m|}A_{Mm}e^{im\psi}
\eea
of the angle $\psi$. The  $k$-dependent part
$ e^{-k\frac{R}{2}}k^{M}$ is non-zero
for $k\neq 0$ and the function $\Theta$
 only vanishes identically
  if $A_{Mm}=0$ for all $m=-M, ... ,M$, i.e. if the $2^M$-pole
field is trivial. 

\subsection{The interaction energy of two scalar multipoles}

The interaction energy of two multipoles 
can be expressed in a remarkably compact way.
Let
\bea
\phi_-(x)=\frac{1}{|x-X_-|}
\eea
be the Coulomb potential centred at $X_-=(0,0,-R/2)$ and consider the
multipole field
\bea
\label{vmultipole}
g_N(x) = \sum_{-N\leq n\leq 0}
B_{Nn} \,\,\partial_3^{N+n} { \partial}^{-n} \phi_-(x) +
\sum_{1\leq n\leq N}
B_{Nn} \,\, \partial_3^{N-n} {\bar \partial}^n \phi_-(x)
\eea
with $B_{N(-n)}=\bar B_{Nn}$.
By the same calculation as for $f_M$ above we find the Fourier transform
of $g_N$ in the $x_1x_2$-plane:
\bea
\label{fourv}
q_N(\bk)=4\pi e^{-k\frac{R}{2}}k^{M}\sum_{-N\leq n \leq N}
\left(\frac{i}{2}\right)^{|n|}B_{Nn}e^{in\psi}.
\eea

The  interaction energy  of the two multipole fields $f_M$ and
$g_N$
\bea
\label{interact}
V_{MN}=\langle f_M,g_N\rangle
\eea
can now be computed using the formula (\ref{inprod}).
Using the factorisation property of the Fourier transforms $p_N$
and $q_N$, it is easy to perform the integration over $\bk$. Assuming without
loss of generality that $M\leq N$ we first carry out  the
integration over the angle $\psi$ to find
\bea
V_{MN}=4\pi \int_0^\infty  dk
 e^{-kR}k^{N+M}\sum_{m=-M}^M 2^{-2|m|} \bar A_{Mm} B_{Nm},
\eea
where we have used the reality condition for the coefficients
$A_{Mm}$ and $B_{Nm}$.
Computing the remaining integral we obtain the final result
\bea
\label{interformula}
V_{MN}= 4\pi\frac{(M+N)!}{R^{M+N+1}} \sum_{m=-M}^{M}
2^{-2|m|} \bar A_{Mm} B_{Nm}.
\eea
This formula has a number of interesting features. The interaction energy
depends only on the separation of the multipoles and
on the combination $\sum_{m=-M}^M2^{-2|m|}\bar A_{Mm} B_{Nm}$ of the
multipole components.
As explained in appendix A, the multipole moments $Q_{Nn}$ of a
$2^N$-pole can be thought of as elements of the $(2N+1)$-dimensional
unitary irreducible representation $W_N$ of $SO(3)$. The vector $B$
with $2N+1$ components
 $B_{Nn}$, $-N\leq n\leq N$ is naturally an element of $W_N$.
Rotations $G\in SO(3)$ about the centre $X_-$ of the multipole field
$g_N$  act  on the multipole components via $B_{Nn} \mapsto
\sum_{n'=-N}^N U^N_{nn'}(G)B_{Nn'}$,
where $U^N$ is a $(2N+1)$-dimensional
irreducible representation  of $SO(3)$ (because of the 
rescaling (\ref{rescone}) and (\ref{resctwo}) 
this is not the standard unitary representation). With our assumption
that $M\leq N$ we can use  the multipole components $A_{Mm}$,
$-M\leq m\leq M$,
to define the  linear form
\bea
\label{multipolemap}
F_A: W_N \rightarrow \RR, \qquad
B\mapsto \sum_{m=-M}^M 2^{-2|m|} \bar A_{Mm} B_{Nm}.
\eea
By assumption, the components $A_{Mm}$ are not all zero, and therefore
the map $F_A$ is non-degenerate.
Writing the formula (\ref{interformula})  in terms of this map as
\bea
\label{interformula2}
V_{MN}= 4\pi\frac{(M+N)!}{R^{M+N+1}}F_A(B) ,
\eea
we immediately deduce the following result.

\begin{theorem}
\label{nicest}
The interaction energy of an $2^M$-pole and a  $2^N$-pole separated
by a distance $R$ is always non-vanishing for some relative orientation
of the two multipoles.  When such an orientation is chosen,
the modulus of the 
 interaction energy decreases with the separation as $R^{-(M+N+1)}$.
\end{theorem}

\noindent {\bf Proof}: Assuming without loss of generality that
the multipoles are separated along the $x_3$-axis and that  $M\leq N$
we have the formula (\ref{interformula2}) for the interaction energy.
Since the map (\ref{multipolemap}) defined in terms
of the  (non-vanishing) multipole components
$A_{Mm}$  of the $2^M$ pole at $X_+$ is non-degenerate it
has a $2N$-dimensional kernel.
It then follows from the irreducibility of the $(2N+1)$-dimensional
representation $W_N$ that there exists a
rotation $G\in SO(3)$ such that $U^N(G)B$ is not in the kernel
of $F_A$ for some $G$. For that $G$ we thus have
$F_A(U^N(G)B)=\kappa \neq 0$ and
$V_{MN}=4\pi\kappa (M+N)!R^{-(M+N+1)}$.\hfill \halmos

\section{Attractive forces  and existence of minima}

The arguments of the previous section apply to
the asymptotic pion fields of the Skyrme solitons
$U^{(1)}$ and  $U^{(2)}$ discussed in  Section 3.
In particular we note that the  interaction energy $\Delta E$
(\ref{key}) for the  leading multipole fields
$u_M$ and $v_N$ is just the sum over iso-components
of pairings of the form (\ref{interact})
\bea
\Delta E_a =  -2\langle u_M^a,v_N^a\rangle \,=\, 2
\int_{x_3=0} dx_1dx_2 \,\,
( u_M^a \partial_3 v_N^a - v_N^a\partial_3 u_M^a).
\eea

Now pick  one of the iso-indices, say $a=1$,  and use iso-rotations
of the Skyrme solitons 
to make sure  that the first iso-components $u^1_M$ and $ v^1_N$
are non-vanishing. 
It then follows from theorem (\ref{nicest}) that we can make
the multipole interaction energy  $\Delta E_1$
non-zero by spatial rotations of one of the Skyrme solitons.
 This fact is the main
 upshot of the calculations in the previous section and a crucial input for
the following argument which was missing in \cite{CK}.

Now consider  the sum
\bea
\Delta E = \Delta E_1+ \Delta E_2 + \Delta E_3.
\eea
We would like to show that we can always arrange for  $\Delta E$
to be negative by a suitable iso-rotation of one of the Skyrme
 solitons.
We may assume  that, possibly after re-labelling the pion fields,
\bea
 \Delta E_1\geq \Delta E_2 \geq \Delta E_3,
\eea
If $\Delta E< 0$ we are done, so suppose that $\Delta E \geq 0$.
Since we know that not all $\Delta E_a$ vanish we
can conclude that $\Delta E_1  > 0$.
Now perform an iso-rotation
of Skyrme soliton  $1$  by $180$ degrees  around the third iso-spin axis.
This reverses the sign of $\pi_1^{(1)}$ and $\pi_2^{(1)}$ and hence
also the sign of $\Delta E_1$  and $\Delta E_2$. The new value of $\Delta E$
is
\bea
\Delta E& =&- \Delta E_1 -\Delta E_2 + \Delta E_3 \nonumber \\
 &=& -\Delta E_1 -(\Delta E_2 - \Delta E_3) < 0
\eea
since $-\Delta E_1 <0$ and, with our ordering,
$-(\Delta E_2 - \Delta E_3)\leq 0$.

Thus, the contribution $\Delta E$ to the interaction energy
of two Skyrme solitons $U^{(1)}$ and $U^{(2)}$
can always be made less than zero
by suitable rotations and iso-rotations of Skyrme soliton 1.
It follows from the discussion at the end of  Section 3
that  for $|N-M|\leq 2$ and  sufficiently large separation parameter $R$,
\bea
\label{tops}
E[U_R] < E[U^{(1)}]+ E[U^{(1)}].
\eea

We conclude with a few comments on the implications of
our result for the question of existence of general Skyrme
solitons. As explained in Section 1, Esteban proved the existence
of Skyrme solitons of arbitrary degree provided the strict inequality
(\ref{inequality}) holds. Our result (\ref{tops}) implies the
inequality in those cases where minima exist in the sectors $l$
and $k-l$, and where the associated multipoles have orders
which do not differ by more than  two.
 Since monopole fields  do not
arise in Skyrme solitons, the interaction energy $\Delta E$ dominates at large
separation  if the leading multipole moments  in Skyrme solitons are
at most octupoles. As explained  at the end of Section 2, the
$B=7$ Skyrme soliton  is believed to have octupoles as leading
multipoles, but there is no numerical evidence for leading
multipoles of higher order. Unfortunately, it seems very difficult
to rule out this possibility in general.

Even if one could prove  (or circumvent)
 the assumption concerning multipoles, the existence of
attractive forces between Skyrme solitons is not sufficient
to establish the inequality (\ref{inequality}) for infima.
Physically it seems reasonable that the existence of attractive
forces should imply the existence of minima in every sector. However,
we have not been able to develop this observation into
 a mathematical proof. Further thoughts and speculations in
this direction can be found in \cite{Schroerss}.

\vspace{1cm}

\vbox{
\noindent{\bf Acknowledgements}

\noindent
BJS acknowledges financial support through an Advanced Research
Fellowship of the Engineering and Physical Sciences Research Council.
MAS thanks  Rafe Mazzeo for directing him to the unique continuation
theorem used in Section  2.
NSM is grateful to Walter Kohn for drawing his attention to the 
formula (\ref{mexp}).
}

\appendix

\section{Spherical Harmonics}
In this appendix we derive the relation between the standard
 spherical harmonics and the
following functions
on $\RR^3-\{0\}$ used in the multipole  expansion in Section 4.1
\bea
\label{Fdefined}
F_{Nn}(x)=\left\{ \begin{array}{ll}  \bar{\partial}^n \partial_3^{N-n}
\left(\frac{1}{r}\right)
& \mbox{if}\quad  n\geq 0 \\{\partial}^{-n} \partial_3^{N+n}
\left(\frac{1}{r}\right)& \mbox{if} \quad n <0
\end{array} \right.
\eea
Here  $N\geq 0$ and $-N\leq n\leq N$ and $\partial =\frac{1}{2}(\partial_1
-i\partial_2)$.
These functions are  harmonic in their domain:
\bea
\label{harmony}
\Delta F_{Nn}= 0.
\eea
They are also homogeneous of degree $-(N+1)$ so that they can be written
as
\bea
F_{Nn}=\frac{1}{r^{N+1}}\Phi_{Nn}(\theta,\varphi),
\eea
where $(\theta,\varphi)$ are the usual spherical coordinates on
the two-sphere centred at the origin.
Since the
Laplace operator takes the following form in spherical coordinates
\bea
\Delta = \frac{1}{r}\frac{\partial^2}{\partial r^2}r + \frac{1}{r^2}
\Delta_\omega,
\eea
where $\Delta_\omega$ is the Laplace operator on the 2-sphere of unit radius,
it follows from (\ref{harmony}) that
\bea
\label{Lsquared}
\Delta_\omega \Phi_{Nn} = -N(N+1)\Phi_{Nn}.
\eea

Define the generator of rotations about the $3$-axis
\bea
J_3 = -i\frac{\partial}{\partial \varphi}
\eea
and express it in terms of complex coordinates $z=x_1+ix_2$
and complex derivatives in the $x_1x_2$ plane:
\bea
J_3 =z\partial - \bar z \bar \partial.
\eea
The operator $\partial $ acts as a raising operator and the
operator $\bar \partial$ acts as a lowering operator for
$J_3$:
\bea
\label{ladder}
[J_3,\bar \partial]=\bar \partial\quad \mbox{and} \quad
[J_3, \partial]= -  \partial.
\eea
Thus if $\phi_n$ is a function on $\RR^3$
which is an eigenfunction of $J_3$ with eigenvalue $n$
then
$\bar \partial  \phi_n$ is an eigenfunction with eigenvalue $n+1$
provided it is not zero. Similarly
$ \partial  \phi_n$ is an eigenfunction  of $J_3$
with eigenvalue $n-1$ provided it is not zero.

It follows from (\ref{Fdefined}) that 
\bea
F_{Nn}= \left\{ \begin{array}{ll}  \bar{\partial}^n F_{N-n,0}
& \mbox{if}\quad n\geq 0 \\ {\partial}^{-n} F_{N+n,0} & \mbox{if}\quad n <0.
\end{array} \right.
\eea
Noting that, by rotational symmetry about the $3$-axis,
\bea
J_3\Phi_{N0} = 0 \quad \mbox{for all} \,\,N,
\eea
we conclude that
\bea
J_3 F_{Nn}=n F_{Nn}.
\eea
It follows that
\bea
\Phi_{Nn}=r^{N+1}F_{Nn}
\eea
also satisfies
\bea
J_3 \Phi_{Nn}=n \Phi_{Nn}.
\eea
Thus, to sum up,  the $\Phi_{Nn}$ are functions on $S^2$,
which are eigenfunctions of both the Laplace operator   (\ref{Lsquared})
and the operator $J_3$ with eigenvalues respectively $-N(N+1)$ and $n$.
It follows from standard harmonic analysis on
$S^2$ that they must be proportional to the spherical harmonics $Y_{Nn}$.

In the case $n=0$ we can
determine the proportionality constant by evaluating both $\Phi_{N0}$
and $Y_{N0}$ on the positive $3$-axis. With the usual normalisation
\cite{Jackson}  we find
\bea
\label{starting}
Y_{N0}=\sqrt{\frac{2N+1}{4\pi}}\frac{(-1)^N}{N!} \Phi_{N0}.
\eea
The relation between $Y_{Nn}$ and $\Phi_{Nn}$ for $n\neq 0$
is harder to compute. Let us assume initially that $n>0$.
Starting with the standard expression
for the associated Legendre function in terms of Legendre
polynomials
\bea
P^n_N(\cos \theta)=(-1)^n\sin^n\theta
\left(\frac{d}{d\cos \theta}\right)^nP_N(\cos \theta)
\eea
and the expression of the spherical harmonic in terms of associated
Legendre function (see e.g. \cite{Jackson} p. 99) we have the relation
\bea
Y_{Nn}(\theta,\varphi)= (-1)^n \sqrt{\frac{(N-n)!}{(N+n)!}}
\left(\sin \theta e^{i\varphi}\right)^n
\left(\frac{\partial}{\partial\cos \theta}\right)^n
Y_{N0}(\theta,\varphi).
\eea
Then using (\ref{starting}) and the definition of $\Phi_{N0}$ we deduce
\bea
\label{moving}
Y_{Nn}(\theta,\varphi)=r^{N+1}\sqrt{\frac{2N+1}{4\pi}}\frac{(-1)^{N+n}}{N!}
\sqrt{\frac{(N-n)!}{(N+n)!}}z^n D_\theta^n\partial_3^N\left(\frac{1}{r}\right),
\eea
where
\bea
D_\theta=\frac{1}{r}\frac{\partial}{\partial \cos\theta} =-\frac{x_3}{\rho}
\frac{\partial}
{\partial \rho} +\partial_3
\eea
and $z=\rho e^{i\varphi}$ with $\rho=r\sin\theta$
as in the main text of the paper.
Then we use the commutation relation
\bea
[D_\theta,\partial_3]=\frac{1}{\rho}
\frac{\partial} {\partial \rho}
\eea
to move $D_\theta$ past $\partial_3$ in (\ref{moving}).
Noting that $D_\theta\, r=0$ we find
\bea
\label{movinger}
Y_{Nn}(\theta,\varphi)=r^{N+1}\sqrt{\frac{2N+1}{4\pi}}
\frac{(-1)^{N+n}}{\sqrt{(N-n)!(N+n)!}}z^n \left(\frac{1}{\rho}
\frac{\partial} {\partial \rho}\right)^n\partial_3^{N-n}
\left(\frac{1}{r}\right).
\eea
Now exploit that on any function $f$ which only depends on $\rho$ and $x_3$
\bea
\frac{1}{\rho}
\frac{\partial f } {\partial \rho} (\rho,x_3)
 = \frac{2}{z}\bar  \partial f(\rho,x_3)
\eea
to conclude
\bea
\label{arriving}
Y_{Nn}(\theta,\varphi)=r^{N+1}\sqrt{\frac{2N+1}{4\pi}}
\frac{(-1)^{N+n}\,2^n}{\sqrt{(N-n)!(N+n)!}}
\bar\partial^n\partial_3^{N-n}\left(\frac{1}{r}\right).
\eea
Thus we finally arrive at promised relationship between  $Y_{Nn}$ and
$\Phi_{Nn}$, valid for $n\geq 0$:
\bea
Y_{Nn}=\sqrt{\frac{2N+1}{4\pi}}\frac{(-1)^{N+n}\,2^n}{\sqrt{(N-n)!(N+n)!}}
\Phi_{Nn}.
\eea
To deduce the corresponding result for $n<0$ we note that $\Phi_{Nn}=
\bar\Phi_{N(-n)}$ and $Y_{Nn}=(-1)^n\bar Y_{N(-n)}$ for $n<0$. Thus
for $n<0$:
\bea
Y_{Nn}=\sqrt{\frac{2N+1}{4\pi}}\frac{(-1)^{N}\,2^{|n|}}{\sqrt{(N-n)!(N+n)!}}
\Phi_{Nn}.
\eea

\end{document}